\documentclass[conference]{IEEEtran}

\usepackage{amsmath}
\usepackage{amsfonts}
\usepackage{amssymb}

\usepackage{amscd}
\usepackage[utf8]{inputenc}
\usepackage{stmaryrd}

\newtheorem{defn}{Definition}[section]
\newtheorem{prop}[defn]{Proposition}
\newtheorem{theo}[defn]{Theorem}

\newtheorem{corolary}[defn]{Corollary}

\newcommand{\citeonline}[1]{\cite{#1}}

\newcommand{\citeauthor}[1]{\cite{#1}}

\title{A Dynamic Preference Logic for reasoning about Agent Programming}

\author{\IEEEauthorblockN{Marlo Souza}
\IEEEauthorblockA{
Federal University of Bahia - UFBA\\
Salvador, Brazil\\
Email: marlo@dcc.ufba.br}
\and
\IEEEauthorblockN{Álvaro Moreira}
\IEEEauthorblockA{
Federal University of Rio Grande do Sul - UFRGS\\
Porto Alegre, Brazil\\
Email: afmoreira@inf.ufrgs.br}
\and
\IEEEauthorblockN{Renata Vieira}
\IEEEauthorblockA{
Pontifical Catholic University of Rio Grande do Sul - PUCRS\\
Porto Alegre, Brazil\\
Email: renata.vieira@pucrs.br}
\and
\IEEEauthorblockN{John-Jules Ch. Meyer}
\IEEEauthorblockA{
Utrecht University\\
Uthrecht, the Netherlands\\
Email: j.j.c.meyer@uu.nl}}

\begin{document}

\maketitle

\begin{abstract}
In this work, we investigate the use of Dynamic Preference Logic to encode BDI mental attitudes. Further, exploring this codification and the representation of preferences over possible worlds by preferences over propositional formulas, here called priority graphs, we comment on how to interpret BDI agent programs in this logic. Also, using the connection between dynamic operations defined over preference models and their encoding as transformations on priority graphs, we show how our logic can be used not only to reason about agent programs, but as a tool to specify reasoning mechanisms to guarantee certain properties in the theory of rationality for the programming language.
\end{abstract}
\begin{IEEEkeywords}
Dynamic Epistemic Logic; Agent Programming; Formal Semantics; BDI Logics;
\end{IEEEkeywords}

\section{Introduction}

The formalisation of mental attitudes have been the object of much discussion in Logic and Philosophy and many such formalisations have been proposed. One of the most influential semantic frameworks in agent specification is the so-called BDI framework, which focuses on the Belief, Desire and Intention attitudes inspiring the development of many agent-oriented programming languages.

While the engineering of such languages has been much discussed, the connections between the theoretical work on Philosophy and Artificial Intelligence and its implementations in programming languages are not so clearly understood yet.  This distance between theory and practice has been acknowledged in the literature for agent programming languages and is commonly known as the ``semantic gap''.

Interpretations for the mental attitudes of the BDI framework have been constructed for some BDI agent-oriented languages such as AgentSpeak(L) \cite{Moreira:Bordini}, GOAL \cite{hindriks2001agent}, etc. These attempts follow the approach of constructing a logic of mental attitudes based on the formal semantics of the programming language. While this approach has the clear advantage of highlighting the meaning of mental attitudes diffused in the programming language, it is not clear how one can use such logics to construct programs - or propose changes in the language semantics - that guarantee certain desired properties.

More yet, one crucial limitation in these attempts to connect agent programming languages and BDI logics, in our opinion, is that the connection is mainly established at the static level, i.e. they show how a given program state can be interpreted as a BDI mental state. Since mental change is not expressible in many of these logics, it is not clear how the execution of a program may be understood as changes in the mental state of the agent. The reason for this, in our opinion, is that the formalisms employed to construct BDI logics are usually static, i.e. cannot represent actions and change,  or can only represent ontic change, not mental change. 

In this work, we will explore the use of Dynamic Preference Logic \cite{girard2008modal}  to encode mental attitudes. 
This logic was shown to have a strong connection with syntactic representations, known as priority graphs \cite{liu2011reasoning}, which can be used as a computational representation of a possible worlds model. We exploit this connection to show how the programming language semantics can be specified by means of mental attitude changes in the corresponding logics and vice-versa.


The structure of this work is as follows:  Section~\ref{sec:dlp} presents Dynamic POreference Logic that will be the foundation of our logic for agents; in Section~\ref{sec:bdi}, we show how Dynamic preference Logic can be used to create a logic for reasoning about an agent's mental state and show how the BDI mental attitudes can be encoded in this logic; in Section~\ref{sec:AAP}, we present some syntactic representations for the models discussed in Sections~\ref{sec:dlp} and \ref{sec:bdi}, and how these representations can be connected to agent programs, as commonly defined in various agent programming languages in the literature. In Section~\ref{sec:rel}, we discuss the related literature and, finally, in Section~\ref{sec:conc}, we present some final considerations about our work.

\section{The Dynamic Logic of Preferences}\label{sec:dlp}

Preference Logic is a modal logic about the class of transitive and reflexive frames. It has been applied to model a plethora of phenomena in Deontic Logic \cite{liu:deontics}, Logics of Preference \cite{boutilier:94}\cite{lang}, Logics of Belief \cite{BAL08} etc. 


Dynamic Preference Logic (DPL) \cite{girard2008modal}\cite{van2009everything} is the result of ``dynamifying'' Preference Logic, i.e. extending it with dynamic modalities - usually represented by programs in Propositional Dynamic Logic (PDL). 

In this section, we introduce Dynamic Preference Logic and some fundamental properties of this logic. This language will be the base for the construction of a logic for BDI reasoning in Section~\ref{sec:bdi}. Let's first introduce the language of (static) Preference Logic.

\begin{defn}
Let $P$ be a finite set of propositional letters. We define the language $\mathcal{L}_{\leq}(P)$ by the following grammar (where $p \in P$): $$
\varphi ::= p ~|~ \neg \varphi ~|~ \varphi \wedge \varphi ~|~ A \varphi ~ | ~ [ \leq ] \varphi~ | ~ [<]\varphi$$
\end{defn}

We will often refer to the language $\mathcal{L}_\leq(P)$ simply as $\mathcal{L}_\leq$, by supposing the set $P$ is fixed. Also, we will denote the language of propositional formulas by $\mathcal{L}_0(P)$ or simply $\mathcal{L}_0$

\begin{defn}
A preference model is a tuple $M = {\langle W, \leq, v\rangle}$ where $W$ is a set of possible worlds, $\leq$ is a reflexive, transitive relation over $W$,  and $v: P \rightarrow 2^W$ a valuation function.
\end{defn}

In such a model, the accessibility relation $\leq$ represents an ordering of the possible worlds according to the preferences of a certain agent. As such, given two possible worlds $w,w' \in W$, we say that $w$ is at least as preferred as $w'$ if, and only if, $w\leq w'$. 

The interpretation of the formulas over these models is defined as usual. 
$$\begin{array}{lll}

M,w \vDash A\varphi &\mbox{ iff } &\forall w' \in W:\, M,w'\vDash\varphi\\
M,w \vDash [\leq]\varphi & \mbox{ iff }&\forall w' \in W:\, w'\leq w \Rightarrow M,w' \vDash  \varphi\\
M,w \vDash [<]\varphi & \mbox{ iff }&\forall w' \in W:\, w'< w \Rightarrow M,w' \vDash \varphi
\end{array}$$
In the definition above, $w<w'$ if, and only if, $w\leq w'$  and $w'\not\leq w$.

As usual, we will refer as ${\langle < \rangle} \varphi$ to the formula $\neg [<]\neg \varphi$. Also, given a model $M$ and a formula $\varphi$, we use the notation $\llbracket \varphi\rrbracket_M$ to denote the set of all the worlds in $M$ satisfying $\varphi$. When it is clear to which model we are referring to, we will denote the same set by $\llbracket \varphi\rrbracket$. 

Given a set of worlds $\llbracket \varphi \rrbracket$ and a (pre-)order $\leq$, we will denote the minimal elements of $\llbracket \varphi \rrbracket$, according to $\leq$, by the notation $Min_\leq \llbracket \varphi \rrbracket$. This corresponds to the notion of `most preferred worlds satisfying $\varphi$' in the model.
This notion can be defined in this logic by the following formula:
\[
\mu\varphi \equiv \varphi \wedge \neg {\langle < \rangle} \varphi
\]

The existence of minimal worlds satisfying $\varphi$ is not always guaranteed, since infinite descending chains may exist in the model. If the relation $<$ in a model $M$ is well-founded, however, we can always guarantee that $\llbracket \mu\varphi \rrbracket_M  = Min_\leq \llbracket \varphi\rrbracket_M \neq \emptyset$. 
A complete axiomatization for the logic restricted to well-founded models has been provided by Souza~\citeonline{souzaphd}.
%

Moreover, Souza et al.~\citeonline{souzakr} showed that if preference models are well-founded some important operations over preference relations, such as some forms of iterated belief revision and contraction,  are well-defined in Preference Logic, expanding the results of Liu~\citeonline{liu2011reasoning}.

\subsection{Dynamics of preferences}\label{sub:dynamics}

In this section, we ``dynamify'' Preference Logic by introducing dynamic modalities representing standard mental change operations such as revisions and contractions. In this work, we present the operations of  public announcement \cite{plaza}, radical upgrade \cite{van2007dynamic} and natural contraction \cite{ramachandran2012three}. The choice for these three operations was motivated by the fact that they are each a representative of a large class of important mental changing operations studied in the literature, namely expansion, revision and contraction, and for the fact that these operations have been well studied in the framework of Dynamic Preference Logic before \cite{pref}\cite{liu2011reasoning}\cite{souzakr}.

The first operation we present is the well-known public announcement of Plaza~\citeonline{plaza}. A public announcement of $\varphi$ is a truthful and knowledge increasing announcement of $\varphi$. 

\begin{defn}\label{def:PA}
\cite{girard2008modal}
Let $M = {\langle W, \leq, v\rangle}$ be a preference model and $\varphi$ a formula of $\mathcal{L}_0$. We say the model  $M_{!\varphi} = {\langle W_{!\varphi}, \leq_{!\varphi},  v_{!\varphi}\rangle}$ is the result of public announcement of $\varphi$ in $M$,  where: 

$
\begin{array}{lll}
W_{!\varphi} &=& \{w \in W ~|~ M,w\vDash \varphi\}\\
\leq_{!\varphi} &=& \leq \cap ~(W_{!\varphi}^2)\\
v_{!\varphi}(p) &=& v(p)~\cap~ W_{!\varphi} 
\end{array}
$%
\end{defn}

%

The radical upgrade of a model by an information $\varphi$ results in a model such that all worlds satisfying $\varphi$ are deemed preferable than those not satisfying it. 

\begin{defn}
\label{def:RU}
Let $M = {\langle W, \leq, v\rangle}$ be a preference model and $\varphi$ a formula of $\mathcal{L}_0$. We say the model  $M_{\Uparrow\varphi} = {{\langle W, \leq_{\Uparrow\varphi},  v\rangle}}$ is the result of the radical upgrade of $M$ by $\varphi$,  where 
$$
\begin{array}{ll}
\leq_{\Uparrow \varphi} = & (\leq \setminus \{{\langle w,w' \rangle} \in W^2~|~ M,w\not\vDash\varphi \mbox{ and }M,w'\vDash\varphi \})\cup\\
&\{{\langle w,w' \rangle} \in W^2~|~ M,w\vDash\varphi \mbox{ and }M,w'\not\vDash\varphi \}
\end{array}$$
\end{defn}


%

Natural contraction is a conservative contraction operation, in the sense that it aims to achieve some form of ``minimal change'' in the belief state. In other words,  the preference relation is changed only in regards to the minimal worlds not satisfying the property to be contracted. 
We define this operation by means of the following transformation on preference models.

\begin{defn}
\label{def:nat}
Let $M = {\langle W, \leq, v\rangle}$ be a preference model and $\varphi$ a formula of $\mathcal{L}_0$. We say the model  $M _{\downarrow\varphi} = {\langle W, \leq _{\downarrow\varphi},  v\rangle}$ is the natural contraction of $M$ by $\varphi$,  where: 
$${\footnotesize
w\leq _{\downarrow\varphi} w'\mbox{ iff }\begin{cases} w\in Min_\leq W \mbox{ or }\\
                                w\in Min_\leq \llbracket\neg \varphi\rrbracket_M \mbox{ or}\\
w \leq w' \mbox{ and } w' \not\in Min_\leq \llbracket \neg \varphi \rrbracket_M\end{cases}
}$$
\end{defn}

%

For each operation $\star$ defined above, we introduce in our language a new modality $[\star\varphi]\psi$ in our language, meaning ``\textit{after the operation of} $\star$ \textit{by} $\varphi$, $\psi$ \textit{holds}''. which can be interpreted as
$$M,w\vDash [\star\varphi]\psi \qquad \mbox{iff} \qquad M_{\star\varphi},w\vDash \psi$$

An important result about Dynamic Preference Logic with these operations is that it has the same expressibility as Preference Logic studied before. In fact, the formulas $[!\varphi]\psi$, $[\Uparrow\varphi]\psi$ and $[\downarrow\varphi]\psi$ are definable in the language of Preference Logic.

\section{A dynamic logic of for BDI agent programming}\label{sec:bdi}

In this section, we use Dynamic Preference Logic to model the mental attitudes of the BDI framework. 
Preference Logic has been used to encode several different mental attitudes in the literature before, among them knowledge, beliefs \cite{BAL08} and goals or desires \cite{boutilier:94}\cite{lang}. In this section, we propose a logic encoding both notions. 
 For that, we introduce two (box) modalities in the language one for encoding the notion of \textit{plausibility}, written $[\leq_P]$, and one for \textit{preference} or \textit{desirability}, written $[\leq_D]$. As such, we construct the language of this logic below. 

\begin{defn}\label{def:lang}
We define the language $\mathcal{L}_{\leq_P,\leq_D}(P)$ by the following grammar (where $p \in P$ a set of propositional letters): $$
\varphi ::= p ~|~ \neg \varphi ~|~ \varphi \wedge \varphi ~|~ A \varphi ~ | ~ [\leq_P]\varphi ~|~ [<_P] \varphi| ~ [\leq_D]\varphi ~|~ [<_D] \varphi$$
\end{defn}

As before, we will define $E\varphi \equiv \neg A \neg \varphi$ and ${\langle \leq_\Box \rangle} \varphi \equiv \neg {[\leq_\Box]}\neg\varphi$ with $\Box \in \{P,D\}$.  The formula $[\leq_D]\varphi$ (${[\leq_P]\varphi}$) means that in all words equally or more desirable (plausible) than the current one, $\varphi$ holds and $[<_D]\varphi$ ($[<_P]\varphi$) that in all words strictly more desirable (plausible) than the current one, $\varphi$ holds.


To interpret these formulas, we will introduce a new kind of Kripke model containing two accessibility relations - one for plausibility and one for desirability. We will call this new model an \textit{agent model}.%

\begin{defn}\label{def:agmodel}
An agent model is a tuple $M = {{\langle W, \leq_P,\leq_D, v\rangle}}$ where $W$ is a set of possible worlds, and both $\leq_D$ and $\leq_P$ are pre-orders over $W$ with well-founded strict parts $<_P$ and $<_D$ and $v$ is a valuation function.
\end{defn}

Notice that an agent model is an amalgamation of two different preference models encoding the orderings for plausibility and desirability. 
%
%
%
%
The interpretation of the formulas is defined as usual, with each modality corresponding to an accessibility relation. We will call $\mu_P \varphi$ ($\mu_D \varphi)$ the formula with the same structure as $\mu\varphi$ when using the modality $<_P$ (resp. $<_D$), i.e. $$\mu_P \varphi \equiv \varphi \wedge \neg {\langle <_P\rangle} \varphi$$

Similar to what was done in Preference logic, we can dynamify our agent logic by including dynamic modalities such as $[\Uparrow_P \varphi]\psi$ to mean that ``\textit{after the radical upgrade of the plausibility relation by} $\varphi$, $\psi$ holds''.

Once we introduced the language we will use in our work, let's encode the notions of mental attitudes. In Philosophical Logic, particularly Deontic Logic, it has been argued that mental attitudes are conditional in nature \cite{hansson:deontics}. These conditional attitudes have been traditionally expressed by means of dyadic modalities of the form $C(\psi | \varphi)$ to represent `\textit{in the context of }$\varphi$, $C\psi$.' Conditionals are common in planning and practical reasoning, being used, for example, to express dependency relations among the agent's desires. In this work, we will encode mental attitudes by conditional modalities.

Let's start with encoding 
%
beliefs. We want to define a conditional modality $B(\psi|\varphi)$ meaning that `\textit{in the context of} $\varphi$, \textit{it is most plausible that}, $\psi$ holds.'  We propose the following codification of conditional belief:

\[
B(\psi|\varphi) \equiv A(\mu_P \varphi\rightarrow\psi)\]

Clearly, the semantics of $B(\psi|\varphi)$ implies that the most plausible $\varphi$-worlds are $\psi$-worlds, i.e. $Min_{\leq_P} \llbracket \varphi\rrbracket \subseteq \llbracket \psi \rrbracket$. 
Finally, we define the unconditional belief $B(\psi)$, meaning `\textit{it is most plausible that} $\psi$ \textit{holds}', as $B(\psi) \equiv B(\psi|\top)$.

Encodings of the notion of desire are numerous in the literature with various meanings according to the intended application. For the sake of our modeling, we will require that agent's desires are consistent with each other - a common requirement in logical modelling of desires. Hindriks et al.~\citeonline{hindriks2001agent} argues that, since desires are future-directed in nature, such restriction is not necessary, for an agent needs not to desire to achieve $\varphi$ and $\neg \varphi$ \textit{at the same time}. We agree with their criticism and point out that the kinds of desires they aim to represent can be expressed in our language as well, but for the aim of encoding intentions  this simple kind of desires will suffice.

Similar to belief, we propose a codification of desires as everything that is satisfied in all most desirable worlds. In other way, we want to encode a formula $G(\psi|\varphi)$ meaning that ``\textit{in the most desirable} $\varphi$\textit{-worlds,} $\psi$ \textit{holds}''. As such, we can encode goals as: 
\[G(\psi|\varphi) \equiv A(\mu_D \varphi \rightarrow \psi)\]

Our encoding of desires is similar to \citeauthor{boutilier:94}'s  ideals in Qualitative Decision Theory. It is our belief that Boutilier's ideals model quite faithfully the notion of \textit{overwhelming desire}, i.e. a desire that is always preferred to its alternatives. As such, the formula $G(\psi)\equiv G(\psi|\top)$ models the fact that the agent `\textit{necessarily wants that} $\psi$', i.e. in the most desirable worlds $\psi$ holds.

There is no consensus on which properties a theory of intentions should satisfy to properly describe the notions of intentional action, intentionality, etc. In the Artificial Intelligence research, Cohen and Levesque's \citeonline{Cohen90} desiderata for intentions based in Bratman's\citeonline{bratman} work  has become the official benchmark for any theory aiming to formalise such notions. 


Central to Cohen and Levesque's requirements, in our understanding, are two distinctive roles of intention in practical reasoning: the role of intention as a constraint in the possible actions/desires entertained by the agent 
and intention as a product of practicality, i.e. intentions as intrinsically connected to plans.

Since our logic does not possess the expressibility to refer to ontic actions, i.e. actions that change the current state of the world, we propose an initial codification of `admissible intention', $AdmInt(\psi|\varphi)$, i.e. a property that satisfies the consistency requirements of Bratman, and may be eventually adopted as a prospective intention. This notion later will be refined, when we enrich the language to include ontic actions. Bratman's \cite{bratman}, simplified by Cohen and Levesque's desiderata, expresses the relationship between the attitudes of intention, desire and belief. Particularly, according to this requirements, an intention is a desire that the agent believes to be possible and that has not yet been achieved. We can model this relation in the following way, where $AdmInt(\psi|\varphi)$ means that `\textit{in the context of} $\varphi$ \textit{it is admissible to intend to achieve} $\psi$ ':
\[
AdmInt(\psi|\varphi)\equiv G(\psi|\varphi)\wedge E(\psi\wedge \varphi)\wedge \neg B(\psi|\varphi)
\]

In the following we will extend our logic to include ontic actions. With that extention, we can propoerly express the relationship between intentions and practical reason, i.e. how intention and actions are connected.

\subsection{Intentions and practicality}
The relationship between intention and practicality is quite a different aspect than what we have been treating before. In our framework we do not have the machinery to represent ontic actions - i.e. actions that change the environment. To allow the representation of practicality, we must extend the language of $\mathcal{L}_{\leq_P,\leq_D}$ to incorporate ontic actions, or simply plans.

\begin{defn}\label{def:ra}
We call $\mathcal{P} = {\langle \Pi, pre, pos\rangle}$ 
an action library, or plan library, iff $\Pi$ is a set finite set of plans symbols, $pre: \Pi \rightarrow \mathcal{L}_0$ is a function that maps each plan to a propositional formula representing its preconditions and $pos: \Pi \rightarrow \mathcal{L}_0$ the function that maps each plan to a propositional formula representing its post-conditions. 
We further require that  the post-conditions of any plan is a consistent conjunction of propositional literals. 
We say $\alpha \in \mathcal{P}$ for any plan symbol $\alpha \in \Pi$.
\end{defn}

To model the effect of performing an ontic action $\alpha \in \mathcal{P}$ given an agent model $M$, we will define the notion of model update, as commonly used in the area of Dynamic Epistemic Logic. 

\begin{defn}
Let $\mathcal{P} = {\langle \Pi,  pre, pos \rangle}$ be a plan library, $\alpha \in \mathcal{P}$ an action (or plan) and $M= {\langle W, \leq_P,\leq_D, v\rangle}$ an agent model. The product update of model $M$ by action $\alpha$ is defined as the model $M \otimes [\mathcal{P},\alpha] = {\langle W', \leq_P', \leq_D', v'\rangle}$ where
$$\begin{array}{lll}
W' & = & \{w\in W ~|~ M, w\vDash pre(\alpha)\}\\
\leq_P' & = & \leq_P \cap~ W'\times W'\\
\leq_D' & = & \leq_D \cap~ W'\times W'\\
v'(p) &=& \begin{cases}
					W' & \mbox{if }pos(\alpha)\vDash p\\
					\emptyset &\mbox{if }pos(\alpha)\vDash \neg p\\
					v(p) \cap~ W'\times W'& \mbox{otherwise}
\end{cases}
\end{array}$$
\end{defn}

Bratman~\citeonline{bratman} defends the thesis that intentions are intrinsically connected to plans, in the sense that intentions are the plans that the agent adopts to achieve a certain desired state of the world. These (procedural) intentions, however, are constrained by a series of consistency requirements, most of which we encoded by means of the formula $AdmInt (\psi|\varphi)$. We define, thus, when a set of plans are considered admissible as the (procedural) intentions of an agent in a given state of mind.

\begin{defn}
Let $\mathcal{P}$ be a plan library and $M = {\langle W, \leq_P,\leq_D, v\rangle}$ be an agent model. We say a set $I\subset \Pi$ of plans is $\mathcal{P}$-consistent with $M$ if for all $\alpha \in I$, $M\vDash B(pre(\alpha))$ and $M\vDash AdmInt(pos(\alpha))$. If $I$ is $\mathcal{P}$-consistent with $M$, we say $M' = {\langle W, \leq_P,\leq_D, I,v\rangle}$ is a practical agent model.
\end{defn}

With that, we can expand our language to include actions and procedural intentions, i.e. formulas of the sort $[\alpha]\varphi$ and $I\alpha$, meaning that `\textit{after the execution of }$\alpha$, $\varphi$ \textit{holds}' and `\textit{it is intended to} $\alpha$,' respectively.

\begin{defn}\label{def:modelplan}
Let $\mathcal{P}$ be a plan library and $M = {\langle W, \leq_P,\leq_D, I, v\rangle}$ be a practical agent model. For any $\alpha \in \mathcal{P}$, we introduce the formulas $[\alpha]\varphi$ and $I\alpha$ and define
$$
\begin{array}{lll}
M,w \vDash [\alpha]\varphi &\mbox{iff} & M,w\vDash pre(\alpha) \Rightarrow M\otimes [\mathcal{P},\alpha],w \vDash \varphi\\
M,w \vDash I\alpha &\mbox{iff} & \alpha \in I\\
\end{array}$$
\end{defn}

With the addition of actions, we can represent the notion of ability. An agent can achieve $\varphi$ if there is an executable action $\alpha$, i.e. $pre(\alpha)$ holds, that causes $\varphi$ to hold. More yet, Bratman requires that Intentions are intrinsically connected to plans, meaning that if an agent intends to achieve a state of affairs, she must have a  plan to achieve it. With these requirements we propose the following codification for intentions:

%

\[
Int (\psi|\varphi) \equiv AdmInt(\psi|\varphi)\wedge \bigvee_{\alpha \in \mathcal{P}} (I\alpha \wedge B\left(pre(\alpha)\wedge [\alpha]\psi~|~ \varphi\right))
\]
Meaning that ``\textit{in the context of} $\varphi$ \textit{the agent intends to achieve} $\psi$''. As before we define $Int(\varphi) \equiv Int(\varphi | \top)$.

It is easy to see by our construction that procedural intentions, i.e. intentions to do, and prospective intentions, i.e. intentions to be, are well-connected.
\begin{prop}
Let $\mathcal{P}$ be a plan library and $M = {\langle W, \leq_P,\leq_D, I, v\rangle}$ be a practical agent model, it holds that
$$M,w \vDash I\alpha \Rightarrow M,w\vDash B(pre(\alpha)) \wedge Int (pos(\alpha))$$
\end{prop}

\section{Agent Logic and Agent Programming}\label{sec:AAP}

Now, we will focus our attention to the connection between our logic and agent programs. 
To understand this connection, we will explore some representation results relating preference models and a syntactic structure to encode preference relations, known as priority graphs. 


\begin{defn} \cite{liu2011reasoning}
Let $\mathcal{L}_0(P)$ be the propositional language constructed over the set of propositional letters $P$, as usual. A P-graph is a tuple $\mathcal{G} = {\langle \Phi, \prec \rangle}$ where $\Phi \subset \mathcal{L}_0(P)$, is a set of propositional sentences and $\prec$ is a strict partial order on $\Phi$.
\end{defn}

A priority graph is a partial order over a set of propositional formulas. The connection between these preferences over formulas and preferences over possible worldshas been studied in the literature \cite{lang}. In our work, following \citeonline{liu2011reasoning},  we use the lexicographic ordering to provide this connection.

\begin{defn}
\label{def:leqg}
\cite{liu2011reasoning}
Let $\mathcal{G} = {\langle \Phi, \prec\rangle}$ be a P-graph, $W$ be a non-empty set of states or possible worlds, and $v: P \rightarrow 2^W$ be a valuation function. The order relation $\leq_\mathcal{G}~\subseteq~ W^2$ is defined as follows: 
$$\begin{array}{ll}
w \leq_\mathcal{G} w'  \mbox{iff} \forall \varphi \in \Phi:& (w' \vDash\varphi \Rightarrow w \vDash\varphi)\vee\\
& (\exists \psi \prec\varphi:(w\vDash \psi \mbox{ and } w'\not\vDash\psi))  \end{array}$$
\end{defn}

From Definition~\ref{def:leqg}, we can say a model $M = {\langle W, \leq_\mathcal{G},v\rangle} $ is induced by a given priority graph $\mathcal{G}$ when its preference relation is constructed as above.

\begin{defn}\label{def:model}
Let $\mathcal{G} = {\langle \Phi, \prec\rangle}$ a P-graph and  $M = {\langle W, \leq, v\rangle}$ a preference model. We say $M$ is induced by $\mathcal{G}$ iff $\leq ~=~ \leq_\mathcal{G}$, where $\leq_\mathcal{G}$ is the relation defined in Definition~\ref{def:leqg} over the set $W$ considering the valuation $v$.
\end{defn}

Liu~\citeonline{liu2011reasoning} shows that any model with a reflexive, transitive relation is  induced by some priority graph.

\begin{theo}
\label{teo:pgraph}
\cite{liu2011reasoning}
Let  $M = {\langle W, R, v\rangle}$ a modal model. The following two statements are equivalent:
\begin{enumerate}
\item $M = {\langle W, R\rangle}$ is a preference frame;
\item There is a priority graph $\mathcal{G}=(\Phi,\prec)$ and a valuation $v$ on $M$ s.t. $\forall w,w' \in W: ~w R w'$ iff $w \leq_\mathcal{G} w'$.
\end{enumerate}
More yet, if $W$ is finite, then so is $\Phi$.
\end{theo}

Since in this work agent models are nothing more that the union of two preference models, we know that there must be a similar syntactic representation for agent models as well. We will define, thus, the notion of an agent structure, which will serve as this syntactic counterpart for agent models.

\begin{defn}
Let $\mathcal{L}_0(P)$ be the propositional language constructed over the set of propositional letters $P$, as usual. An agent structure is a pair $\mathcal{G} ={\langle \mathcal{G}_P, \mathcal{G}_D\rangle}$, where both $\mathcal{G}_P = {\langle \Phi_P, \prec_P\rangle}$ and $\mathcal{G}_D = {\langle \Phi_D, \prec_D\rangle}$ are P-graphs.
\end{defn}

From agent structures we define the notion of induced agent model, similar to what was done to preference models in Definition~\ref{def:model}. We just need to take the P-graphs that induce the plausibility and desirability relations ($\leq_P$ and $\leq_D$, respectively) which are guaranteed to exist by Theorem~\ref{teo:pgraph}.

\begin{defn}\label{def:agentstructure}
Let $\mathcal{G} = {\langle \mathcal{G}_P, \mathcal{G}_D\rangle}$ be an agent structure and $M = {\langle W, \leq_P, \leq_D, v\rangle}$ an agent model. We say $M$ is induced by $\mathcal{G}$ iff $\leq_P \,=\, \leq_{\mathcal{G}_P}$ and $\leq_D \,=\, \leq_{\mathcal{G}_D}$.
\end{defn}

From Definition~\ref{def:agentstructure}, it is clear that every agent model is induced by some agent structure. 

\begin{corolary}\label{teo:agentstructure}
Let $M={\langle W, \leq_P, \leq_D, v\rangle}$ be an agent model. There is an agent structure  $\mathcal{G}={\langle\mathcal{G}_P,\mathcal{G}_D\rangle}$ s.t. $M$ is induced by $\mathcal{G}$.
\end{corolary}

In most BDI agent programming languages, an agent program is defined by means of a tuple $ag = {\langle K, B, D, I\rangle}$, where $K$, $B$ and $G$ are sets of (ranked) propositional formulas representing the agent's knowledge, beliefs and desires, respectively, and $I$ is a set of plans adopted by the agent. Since a set of (ranked) formulas is nothing more than an order over formulas, we can construct an agent structure $\mathcal{G}$ which induces an agent model $M_\mathcal{G}$ representing the mental state of the agent program $ag$.

\begin{defn} \label{def:agentprogram}
Let $ag = {\langle K, B, D, I\rangle}$ be a tuple, where $K$ is a consistent set of propositional formulas, $B$ and $D$ are priority graphs and $I$ is a set of actions in an action library $\mathcal{P}$. We define the model induced by $ag$ as $M= {\langle \llbracket K\rrbracket, \leq_B, \leq_D, v\rangle}$ where $\llbracket K \rrbracket \subset 2^P$ are all the propositional valuations that satisfy the set $K$, $\leq_B\subset \llbracket K \rrbracket \times \llbracket K \rrbracket$ and $\leq_D\subseteq \llbracket K \rrbracket \times \llbracket K \rrbracket$ are the preference relations induced by the graphs $B$ and $D$, and $w\in v(p)$ iff $p\in w$. If $I$ is $\mathcal{P}$-consistent with $M$, then $M_{ag} = {\langle\llbracket K\rrbracket, \leq_B, \leq_D, I, v\rangle}$ is the practical agent model induced by $ag$.
\end{defn}

In Section~\ref{sec:dlp}, we introduced three dynamic operations in the logic of Preference Logic. In our agent logic these operations gain an interpretation of mental change, based on the results in the agent's mental attitudes. As such, public announcements can be understood as knowledge acquisition, while radical upgrade and natural contraction can be understood as either belief revision/contraction or preference revision/contraction. 

\begin{theo}[\cite{souzaphd}]
All the dynamic operations presented in Subsection~\ref{sub:dynamics} are definable by means of operations in P-graphs, if we consider induced models defined in Definition~\ref{def:agentprogram}.
\end{theo}

Since the semantics of agent programming languages (and deliberation mechanisms) can be specified by means of the transformation on the agent's mental state, if we can specify a desirable property one wishes the programming language semantics (or deliberation mechanism) to satisfy by means of these actions, we can automatically generate the corresponding semantic rule by means of transformation of agent programs, using the established correspondence between operations on preference models and operations in agent structures. For example, if one wishes to implement a belief revision such that every time an agent comes to believe $\varphi$ she will drop her intentions to $\neg \varphi$. We can define such operation, let's call it $M_{\uparrow \varphi}$ as a composition of the operations of preference contraction and belief revision $(M_{\downarrow_D \neg \varphi})_{\Uparrow_P \varphi}$, which can be translated as an operation in priority graphs.

\section{Related Work}
\label{sec:rel}
From the Agent Programming perspective, the two most important works on modelling BDI mental attitudes are, in our opinion, the seminal work of Cohen and Levesque~\citeonline{Cohen90} and the work of Rao and Georgeff~\citeonline{RaoGeorgeff} describing the logic BDI-CTL. While their contribution to the area is undeniable, much criticism has been drawn to both approaches. Particularly, both approaches have proven to be difficult to connect with agent programming languages, by the use of a possible-world model semantics - vastly different from the syntactical representations used in agent programming.

Other work have also been proposed for studying the declarative interpretation of mental attitudes in concrete agent programming languages. Works as that of Wobcke \citeonline{wobcke2004model} and of Hindriks and Van der Hoek \citeonline{hindriks2008goal} propose ways to connect the semantics of a given programming language to some appropriate logic to reason about agent's mental attitudes. While they are important in allowing us to analyse the mental attitudes diffused in the semantics of the language, since these logics cannot represent mental actions, the transformations in the agent program, which are defined in the programming language semantics, cannot be understood within the logic used to analyse these mental attitudes and thus the dynamic properties of these attitudes cannot be reasoned about in the logic. Also, in this approach, it is not clear how to establish the contrary connection, i.e. how to create or change programs to guarantee a certain property in the theory of intentions. In our work, since we can translate both ways, from the logic to agent programs and back, this is not an issue.


On the other way, works as that of Bordini and Moreira~\citeonline{Moreira:Bordini} present a declarative interpretation of BDI attitudes based on the actual implementation of these concepts in a concrete agent programming language. The aim of their work is to analyse Rao and Georgeff's \citeonline{RaoGeorgeff} asymmetry properties in the formal semantics of the language AgentSpeak(L). The result is that, due to implementation considerations of the programming language, the logic suffers from a great expressibility limitation, not being able to represent several important properties about mental states.


Perhaps the work most related to ours in spirit is that of \citeonline{hindriks2009toward}. They propose a dynamic logic for agents and show that this logic can be understood as a ve\-ri\-fi\-ca\-tion logic, i.e. it has an equivalent state-based semantics based on the an  operational semantics. The main difference of their approach to ours is that the authors choose to work in a framework closely related to situation calculus. The mental actions involved in decision making and in mental change are, thus, only implicitly defined, while the inclusion of such actions in the language is exactly the main advantage advocated by us.  In some sense, our work can be seen as a generalisation of their work, since by employing Dynamic Preference Logic the equivalence they seek between operational semantics and declarative semantics can be automatically achieved by the results of Liu~\citeonline{liu2011reasoning}.

Recently, Herzig et al. \citeonline{herzig16} pointed out some deficiencies in the formal frameworks for specifying BDI agents which are available in the literature. The authors point out the advantages of a formal theory with a close relationship with the work in belief dynamics and with agent programming. 

\section{Final Considerations}\label{sec:conc}

Our work has proposed a logic for reasoning about BDI agents and a connection between this logic and agent programs, as usually described in agent programming languages. Our logic is flexible enough to specify different mechanisms for agent deliberation as well as different properties for beliefs, desires and intention from the codifications proposed in this work. As such, we believe this logic to be applicable to reason about programs for many agent programming languages.
  
Regarding the requirements proposed by Herzig et al.~\citeonline{herzig16} for a formal theory of agent programming, we believe our work tackles most of the problems identified by those authors. It remains, however, to provide a greater connection of our logics with the work areas as \textit{planning} and \textit{game theory}. We point out, however, that we have powerful evidences that such connections can be done. For example, the work of Andersen et al.~\citeonline{andersen2014don} explore how to integrate planning in the dynamic logics as the one we propose. For the connection with decision theory and game theory, we point out that utilitarian interpretations of agent models have been provided by Boutilier~\citeonline{boutilier:94}. Also, the work of Roy~\citeonline{roy} provides codification of intentions in epistemic game theory using possible worlds models related to ours in which each possible world is a strategy. We believe we can provide a connection between our agent models and Roy's semantics without many difficulties.


\bibliographystyle{IEEEtran}
\bibliography{ijcai17}

\begin{thebibliography}{10}
\providecommand{\url}[1]{#1}
\csname url@samestyle\endcsname
\providecommand{\newblock}{\relax}
\providecommand{\bibinfo}[2]{#2}
\providecommand{\BIBentrySTDinterwordspacing}{\spaceskip=0pt\relax}
\providecommand{\BIBentryALTinterwordstretchfactor}{4}
\providecommand{\BIBentryALTinterwordspacing}{\spaceskip=\fontdimen2\font plus
\BIBentryALTinterwordstretchfactor\fontdimen3\font minus
  \fontdimen4\font\relax}
\providecommand{\BIBforeignlanguage}[2]{{%
\expandafter\ifx\csname l@#1\endcsname\relax
\typeout{** WARNING: IEEEtran.bst: No hyphenation pattern has been}%
\typeout{** loaded for the language `#1'. Using the pattern for}%
\typeout{** the default language instead.}%
\else
\language=\csname l@#1\endcsname
\fi
#2}}
\providecommand{\BIBdecl}{\relax}
\BIBdecl

\bibitem{Moreira:Bordini}
R.~Bordini and A.~Moreira, ``Proving {BDI} properties of agent-oriented
  programming languages: The asymmetry thesis principles in {A}gent{S}peak
  ({L}),'' \emph{Annals of Mathematics and Artificial Intelligence}, vol.~42,
  no.~1, pp. 197--226, 2004.

\bibitem{hindriks2001agent}
K.~V. Hindriks, F.~S. De{ }Boer, W.~Van{ }der{ }Hoek, and J.-J.~C. Meyer,
  ``Agent programming with declarative goals,'' in \emph{Intelligent Agents VII
  Agent Theories Architectures and Languages}.\hskip 1em plus 0.5em minus
  0.4em\relax New York, US: Springer, 2001, pp. 228--243.

\bibitem{girard2008modal}
P.~Girard, ``Modal logic for belief and preference change,'' Ph.D.
  dissertation, Stanford University, 2008.

\bibitem{liu2011reasoning}
F.~Liu, \emph{Reasoning about preference dynamics}.\hskip 1em plus 0.5em minus
  0.4em\relax New York, US: Springer, 2011, vol. 354.

\bibitem{liu:deontics}
J.~Van{ }Benthem, D.~Grossi, and F.~Liu, ``Priority structures in deontic
  logic,'' \emph{Theoria}, vol.~80, no.~2, pp. 116--152, 2014.

\bibitem{boutilier:94}
C.~Boutilier, ``Toward a logic for qualitative decision theory,'' in
  \emph{Proceedings of the 4th International Conference on Principles of
  Knowledge Representation and Reasoning}.\hskip 1em plus 0.5em minus
  0.4em\relax New York, US: Morgan Kaufmann, 1994, pp. 75--86.

\bibitem{lang}
J.~Lang, L.~Van{ }der{ }Torre, and E.~Weydert, ``Hidden uncertainty in the
  logical representation of desires,'' in \emph{Proceedings of the 18th
  international joint conference on Artificial intelligence}.\hskip 1em plus
  0.5em minus 0.4em\relax New York, US: Morgan Kaufmann, 2003, pp. 685--690.

\bibitem{BAL08}
A.~Baltag and S.~Smets, ``A qualitative theory of dynamic interactive belief
  revision,'' \emph{Texts in logic and games}, vol.~3, pp. 9--58, 2008.

\bibitem{van2009everything}
J.~Van{ }Benthem, P.~Girard, and O.~Roy, ``Everything else being equal: A modal
  logic for ceteris paribus preferences,'' \emph{Journal of philosophical
  logic}, vol.~38, no.~1, pp. 83--125, 2009.

\bibitem{souzaphd}
M.~Souza, ``Choices that make you change your mind: a dynamic epistemic logic
  approach to the semantics of bdi agent programming languages,'' Ph.D.
  dissertation, Universidade Federal do Rio Grande do Sul, 2016.

\bibitem{souzakr}
M.~Souza, A.~Moreira, R.~Vieira, and J.-J.~C. Meyer, ``Preference and
  priorities: A study based on contrction,'' in \emph{Proceedings of the
  Fifteenth International Conference on Principles of Knowledge Representation
  and Reasoning}.\hskip 1em plus 0.5em minus 0.4em\relax AAAI Press, 2016, pp.
  155--164.

\bibitem{plaza}
J.~Plaza, ``Logics of public communications,'' \emph{Synthese}, vol. 158,
  no.~2, pp. 165--179, 2007.

\bibitem{van2007dynamic}
J.~Van{ }Benthem, ``Dynamic logic for belief revision,'' \emph{Journal of
  Applied Non-Classical Logics}, vol.~17, no.~2, pp. 129--155, 2007.

\bibitem{ramachandran2012three}
R.~Ramachandran, A.~C. Nayak, and M.~A. Orgun, ``Three approaches to iterated
  belief contraction,'' \emph{Journal of philosophical logic}, vol.~41, no.~1,
  pp. 115--142, 2012.

\bibitem{pref}
J.~Van{ }Benthem and F.~Liu, ``Dynamic logic of preference upgrade,''
  \emph{Journal of Applied Non-Classical Logics}, vol.~17, no.~2, pp. 157--182,
  2007.

\bibitem{hansson:deontics}
B.~Hansson, ``An analysis of some deontic logics,'' \emph{Nous}, pp. 373--398,
  1969.

\bibitem{Cohen90}
P.~R. Cohen and H.~J. Levesque, ``Intention is choice with commitment,''
  \emph{Artificial Intelligence}, vol.~42, no. 2-3, pp. 213--261, 1990.

\bibitem{bratman}
M.~E. Bratman, \emph{Intention, plans, and practical reason}.\hskip 1em plus
  0.5em minus 0.4em\relax Cambridge, US: Harvard University Press, 1999.

\bibitem{RaoGeorgeff}
A.~S. Rao and M.~P. Georgeff, ``Decision procedures for {BDI} logics,''
  \emph{Journal of Logic and Computation}, vol.~8, no.~3, pp. 293--343, 1998.

\bibitem{wobcke2004model}
W.~Wobcke, ``Model theory for {PRS}-like agents: Modelling belief update and
  action attempts,'' in \emph{Proceedings of the 8th Pacific Rim International
  Conference on Artificial Intelligence}.\hskip 1em plus 0.5em minus
  0.4em\relax Berlin, DE: Springer-Verlag, 2004, pp. 595--604.

\bibitem{hindriks2008goal}
K.~Hindriks and W.~Van{ }der{ }Hoek, ``Goal agents instantiate intention
  logic,'' in \emph{Logics in Artificial Intelligence}.\hskip 1em plus 0.5em
  minus 0.4em\relax New York, US: Springer, 2008, pp. 232--244.

\bibitem{hindriks2009toward}
K.~V. Hindriks and J.-J.~C. Meyer, ``Toward a programming theory for rational
  agents,'' \emph{Autonomous Agents and Multi-Agent Systems}, vol.~19, no.~1,
  pp. 4--29, 2009.

\bibitem{herzig16}
A.~Herzig, E.~Lorini, L.~Perrussel, and Z.~Xiao, ``{BDI} logics for {BDI}
  architectures: old problems, new perspectives,'' \emph{KI-K{\"u}nstliche
  Intelligenz}, pp. 1--11, 2016.

\bibitem{andersen2014don}
M.~B. Andersen, T.~Bolander, and M.~H. Jensen, ``Don’t plan for the
  unexpected: Planning based on plausibility models,'' \emph{Logique et
  Analyse}, vol.~1, no.~1, 2014.

\bibitem{roy}
O.~Roy, ``A dynamic-epistemic hybrid logic for intentions and information
  changes in strategic games,'' \emph{Synthese}, vol. 171, no.~2, pp. 291--320,
  2009.

\end{thebibliography}

\end{document}